\documentclass[aps,prd,twocolumn,nofootinbib,showpacs,groupedaddress,preprintnumbers,amsmath,amssymb,floatfix]{revtex4-2}

\usepackage[bookmarks, linktocpage, colorlinks = true, linkcolor = blue, urlcolor  = blue, citecolor = blue, anchorcolor = green, hyperindex = true, hyperfigures]{hyperref}

\usepackage{graphicx} 
\usepackage{amssymb}
\usepackage{dcolumn}
\usepackage{multirow}
\usepackage{amsmath,amssymb}
\usepackage{slashed}
\usepackage[usenames]{color}
\usepackage{float}

\newcommand{\MeV}{\text{MeV}}
\newcommand{\GeV}{\text{GeV}}
\newcommand{\fm}{\text{fm}}
\newcommand{\muB}{\mu_{B}}
\newcommand{\muI}{\mu_{I}}
\newcommand{\nB}{n_{B}}
\newcommand{\nsat}{n_{\mathrm{sat}}}

\newcommand{\Nc}{N_{c}}
\newcommand{\NfV}{N_{f}^{\mathrm{V}}}
\newcommand{\NfM}{N_{f}^{\mathrm{M}}}
\newcommand{\calB}{\mathcal{B}}
\newcommand{\calO}{\mathcal{O}}

\newcommand{\Pid}{P_{\mathrm{id}}}
\newcommand{\Lambdabar}{\bar{\Lambda}}

\begin{document}

\preprint{INT-PUB-24-042}
\title{Interplay between the weak-coupling results and the lattice data in dense QCD}
\author{Yuki~Fujimoto}
\email{yfuji@uw.edu}
\affiliation{Institute for Nuclear Theory, University of Washington, Box 351550, Seattle, WA 98195, USA}

\date{\today}

\begin{abstract}
We discuss the interplay between two first-principles calculations of QCD at high density: perturbative results in the weak-coupling regime and the recent lattice-QCD result at finite isospin density.
By comparing these two results, we verify empirically that the weak-coupling calculations of the bulk thermodynamics and the gap parameter for Cooper pairing between quarks can be applicable down to the quark chemical potential $\mu \sim 1\,\GeV$.
Having verified the validity of the weak-coupling results in QCD at finite isospin density, we discuss possible effects on QCD at finite baryon density, which is relevant for the application to realistic environments such as neutron stars, by using the fact that QCD at finite baryon and isospin density have the common weak-coupling expansions.
First, we show the size of the color-superconducting gap at finite baryon density is as small as a few MeV at $\mu = 1\,\GeV$, which implies that the color-flavor locked phase may be unstable against unpairing up to $\mu \sim 1.4\,\text{GeV}$ even in the weak-coupling regime.
We also introduce a prescription to reduce the ambiguity arising from the undetermined renormalization scale in the weak-coupling calculation by matching with the lattice-QCD data.
We demonstrate the effect of such reduction on neutron-star phenomenology by performing the Bayesian analysis.
\end{abstract}

\maketitle

\section{Introduction}

Identifying the inner core composition of neutron stars (NSs) and determining the equation of state (EoS) are vital challenges in nuclear theory.
It has been hypothesized for a long time that the inner core may contain quark matter~\cite{Ivanenko:1965dg, Itoh:1970uw, Collins:1974ky, Baym:1976yu, Freedman:1977gz, Alford:2004pf, Alford:2006vz, McLerran:2007qj, Baym:2017whm, McLerran:2018hbz, Annala:2019puf, Fujimoto:2022ohj, Marczenko:2022jhl, Annala:2023cwx}.
Such quark matter may exhibit interesting phenomena such as color superconductivity~\cite{Barrois:1977xd, Bailin:1983bm, Alford:1997zt, Rapp:1997zu, Alford:2007xm} and color-flavor locking (CFL)~\cite{Alford:1998mk}.

In principle, the above problem can be solved systematically if the EoS can be computed from first principles of Quantum Chromodynamics (QCD), but the means of computation are limited at finite baryon chemical potential $\muB$ or density $n_B$.
Lattice QCD (LQCD) is until now the most successful first-principles approach to nonperturbative QCD, but QCD at finite $\mu_B$ (QCD$_B$) is beyond the reach because of the sign problem~\cite{Troyer:2004ge, Nagata:2021ugx}.

The only available method in QCD$_B$ is perturbation theory.
Namely, one can perform perturbative expansion in chiral effective field theory ($\chi$EFT) at low $\nB$ around the nuclear saturation density $\nsat \simeq 0.16\,\fm^{-3}$~\cite{Drischler:2021kxf}, and use perturbative QCD (pQCD) at high $\nB \gtrsim 40 \nsat$~\cite{Vuorinen:2024qws}.
Those $\chi$EFT and pQCD calculations can be useful inputs when constraining the EoS from the astrophysical data of NSs~\cite{Hebeler:2013nza, Kurkela:2014vha, Annala:2017llu, Tews:2018kmu, Annala:2019puf, Drischler:2020fvz, Huth:2021bsp, Altiparmak:2022bke, Gorda:2022jvk,Somasundaram:2022ztm, Brandes:2022nxa, Jiang:2022tps, Annala:2023cwx,  Brandes:2023hma, Gorda:2023usm,Koehn:2024ape, Koehn:2024set}.

The pQCD results generally have predictive power as long as the QCD coupling constant $\alpha_s$ stays small, but there are issues related to their applicability and the renormalization scale $\Lambdabar$.
Although there is no benchmark for their applicable limit, it is customary to use the pQCD results down to $\muB = 2.6\,\GeV$ for the input to the NS EoS~\cite{Kurkela:2014vha, Annala:2019puf}.
Meanwhile, some authors claim that the color-superconducting gap derived in pQCD is only applicable at $\muB \gtrsim 10^5\,\GeV$~\cite{Rajagopal:2000rs}.
Also, although the pQCD results are free from fine-tuning of parameters, they have the ambiguity arising from $\Lambdabar$, which leads to uncertainties in physical quantities.
In principle, physical quantities should not depend on $\Lambdabar$, but truncating a perturbative series at finite order gives rise to the $\Lambdabar$ dependence, so such uncertainty can be regarded as that of missing higher-order terms beyond the truncated order.

QCD at finite isospin chemical potential $\muI$ (QCD$_I$), in which quark chemical potentials $\mu$ for the degenerate $u$ and $d$ flavors are set to opposite values $\mu_u= \mu_I /2$ and $\mu_d = - \mu_I /2$, is free from the sign problem and thus amenable to lattice computations.
QCD$_I$ has been intensively studied over the decades~\cite{Kogut:2002tm,Kogut:2002zg,Kogut:2004zg,Beane:2007qr,Beane:2007es,deForcrand:2007uz,Detmold:2008gh,Detmold:2008fn,Detmold:2008yn,Detmold:2008bw,Detmold:2010au,Cea:2012ev,Detmold:2012wc,Detmold:2012pi,Endrodi:2014lja,Brandt:2017oyy,Brandt:2018omg,Brandt:2022hwy,Brandt:2023kev,Abbott:2023coj,Abbott:2024vhj}.
Recently, lattice calculations in QCD$_I$ were able to construct states with a substantial number of pions and determine the EoS in the continuum limit up to $\mu \sim 1.6\,\GeV$, which is large enough to be in the weak-coupling regime~\cite{Abbott:2023coj, Abbott:2024vhj}.

Now the question is how we can avail of this nonperturbative lattice data in QCD$_I$ to reveal the properties of matter in QCD$_B$.
One way is to rigorously relate the EoS of full QCD$_B$ to that in QCD$_I$ through the QCD inequality~\cite{Cohen:2003ut}.
This works despite the difference in their ground states because QCD$_I$ is a phase-quenched theory of QCD$_B$.
Recently, a robust bound on the EoS was derived from the LQCD$_I$ data and QCD inequality~\cite{Fujimoto:2023unl, Abbott:2024vhj}.

In this paper, we show another way to utilize the lattice data in QCD$_I$ by comparing them to the weak-coupling formulas.
This is based on the observation that QCD$_B$ and QCD$_I$ have the same perturbative expansion at the quark chemical potential $\mu = \muB / 3 = |\muI| / 2$.
By an explicit comparison, we first verify the validity of the weak-coupling formulas and examine their applicable limit.
We then introduce a new prescription to mitigate the ambiguity in $\Lambdabar$ and put them onto the same footing as the lattice uncertainty.
We finally show the possible impact of the LQCD$_I$ data on the QCD phase diagram, color superconductivity, and the NS EoS.

\section{Weak-coupling results in dense QCD}

Here, we summarize weak-coupling formulas obtained in pQCD at high $\mu$.
We work in $\Nc = 3$ unless otherwise noted.
We write the vector of chemical potential in the flavor basis as $\boldsymbol{\mu} = (\mu_u, \mu_d, \mu_s)$.
We use the value $\boldsymbol{\mu} = (\mu, - \mu, 0)$ for QCD$_I$ to compare with the lattice data, and $\boldsymbol{\mu} = (\mu, \mu, \mu)$ for QCD$_B$.
The number of flavors in the matter that have nonzero chemical potential is denoted as $\NfM$.
The number of flavors in the vacuum is $\NfV=3$.

The perturbative expansion of pressure $P$ up to next-to-next-to-leading order (NNLO) renormalized in the $\overline{\rm MS}$ scheme is~\cite{Freedman:1976xs, *Freedman:1976dm, *Freedman:1976ub, Baluni:1977ms, Fraga:2001id, Vuorinen:2003fs, Kurkela:2009gj}
\begin{align}
    \frac{P_{\rm pQCD}}{\Pid} &= 1 - 2\frac{\alpha_s}{\pi} - \Big[ \NfM \ln \left(\NfM\frac{\alpha_s}{\pi}\right)
    + \frac{\beta_0}{2} \ln \frac{\Lambdabar^2}{(2\mu)^2}\notag\\
    & \quad + 18 - \frac{22}{9}\NfV + 0.22429 \NfM \Big] \left( \frac{\alpha_s}{\pi} \right)^2 + \calO(\alpha_s^3)\,,
    \label{eq:ppqcd}
\end{align}
where $\beta_0 \equiv 11 - (2/3) \NfV$ is the first coefficient of the QCD $\beta$ function.
The pressure of the ideal quark gas $\Pid$ is defined as $\Pid \equiv \NfM \mu^4/(4 \pi^2)$.

The Cooper pairing gap in the $J=0$ channel is~\cite{Son:1998uk, Hong:1999fh, Schafer:1999jg, Pisarski:1999bf, *Pisarski:1999tv, Brown:1999aq, *Brown:1999yd, *Brown:2000eh, Hsu:1999mp, Wang:2001aq}
\begin{align}
    \ln\left(\frac{\Delta}{\mu}\right) =& - \frac{\sqrt{3}\pi}{2\sqrt{c_R}} {\left(\frac{\alpha_s}{\pi}\right)}^{\!\!-\frac12} \!\!\!- \frac52 \ln \bigg[\NfM \frac{\alpha_s}{\pi} \bigg] \notag \\
    &\quad +\ln \frac{2^{\frac{13}{2}}}{\pi} 
     - \frac{\pi^2 + 4}{12 c_R}- \zeta + \calO(\alpha_s^{\frac12})\,,
    \label{eq:gap}
\end{align}
where $\zeta$ is a pairing-specific coefficient; $\zeta = \frac13 \ln 2$ for the CFL phase and $\zeta = 0$ otherwise.
The color factor is $c_R = 2/3$ for $\boldsymbol{\bar{3}}$ diquark channel and $c_R = 4/3$ for singlet $q\bar{q}$ channel~\cite{Son:2000xc, Cohen:2015soa,Fujimoto:2023mvc}.
The pairing gap $\Delta$ contributes to the thermodynamics as a condensation energy density $\delta P$.
This is measured by the difference between the pressure of the quasiparticle vacuum with and without pairing.
The NLO expression of $\delta P$, recently calculated in Ref.~\cite{Fujimoto:2023mvc} by the present author, is
\begin{equation}
    \delta P = d \frac{\mu^2}{4\pi^2} \Delta^2 \left[1 + \frac{2 \pi}{3^{\frac32}\sqrt{c_R}} \left(\frac{\alpha_s}{\pi}\right)^{\frac12} + \calO(\alpha_s)\right]\,,
    \label{eq:pcond}
\end{equation}
where $d$ is a degeneracy factor dependent on the color-flavor structure of the condensate; $d = 12$ for the CFL phase in QCD$_B$, and $d=6$ for QCD$_I$.
As is classified in Appendix~\ref{sec:corr}, $\delta P$ is the only large correction to $P_{\rm pQCD}$ in the weak-coupling regime of QCD$_I$.

For the running coupling constant $\alpha_s$, we use the 5-loop formula renormalized in $\overline{\mathrm{MS}}$ scheme~\cite{Baikov:2016tgj,Luthe:2016ima,Herzog:2017ohr} although the 1-loop expression is sufficient to renormalize the theory; we do so because we fix the $\overline{\rm MS}$ scale to the value $\Lambda_{\overline{\rm MS}} \simeq 338\,\MeV$ read out from the $N_f=3$ LQCD data using the 5-loop formula~\cite{McNeile:2010ji, Chakraborty:2014aca, Ayala:2020odx, Bazavov:2019qoo, Cali:2020hrj, Bruno:2017gxd, PACS-CS:2009zxm, Maltman:2008bx, FlavourLatticeAveragingGroupFLAG:2021npn}.
Below, we do not consider the uncertainty of $\Lambda_{\overline{\mathrm{MS}}}$ because it appears in theory as the ratio $\Lambdabar/ \Lambda_{\overline{\mathrm{MS}}}$, and thus it can be absorbed into the uncertainty of $\Lambdabar$.

The important property we exploit in this work is that these weak-coupling formulas are universal for QCD$_B$ and QCD$_I$.
It is shown that the expression of $P_{\rm pQCD}$ is the same for QCD$_I$ and QCD$_B$ up to the NNLO, and starts to deviate only at N3LO~\cite{Moore:2023glb}; the deviation is negligible, $\sim 1\%$ at $\mu \sim 1\,\GeV$~\cite{Navarrete:2024zgz}.
Also, the expression of $\Delta$ and $\delta P$ is also the same apart from the color factor.
Furthermore, $\alpha_s$ used in both theories are the same since it is sensitive only to $\NfV$ in the vacuum;  the LQCD uses the $\NfV =3$ configuration~\cite{Abbott:2023coj, Abbott:2024vhj} and NSs are also the $\NfV = 3$ environments.
Below, we use the LQCD$_I$ data as a benchmark for the weak-coupling formulas so that we can use those lattice-informed formulas in QCD$_B$ by resorting to the QCD$_B$-QCD$_I$ universality.

\section{Comparison with LQCD}
\begin{figure}
    \centering
    \includegraphics[width=0.95\columnwidth]{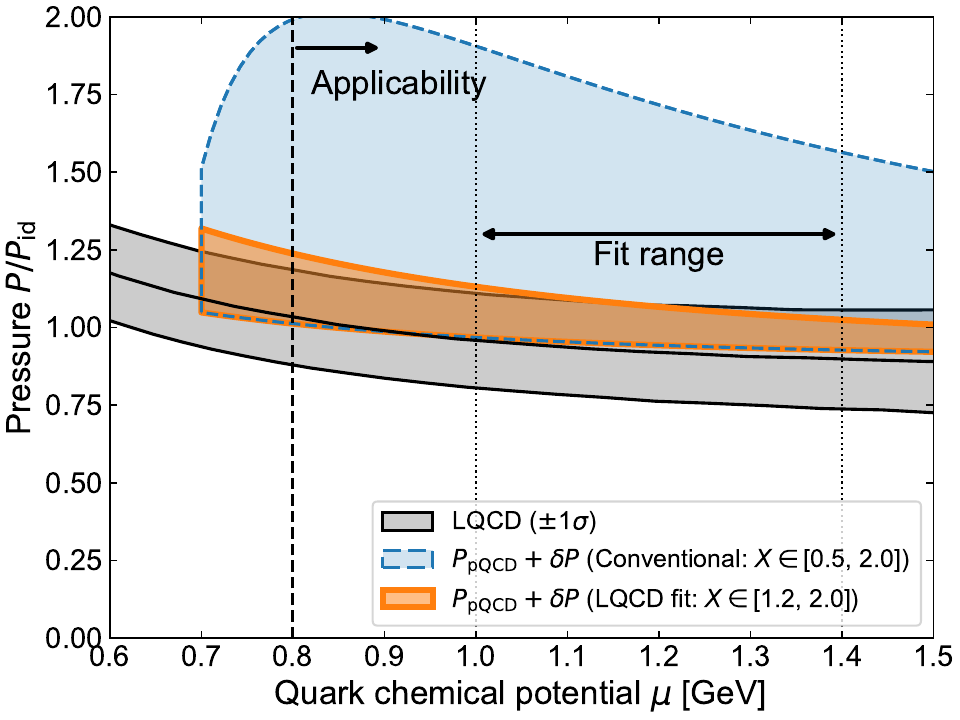}
    \caption{Comparison of the pressure of QCD$_I$ matter normalized by $\Pid$ in the weak-coupling regime and the LQCD data.  The blue band is the weak-coupling pressure with the Cooper pairing effect.  This band is obtained by varying the scale parameter $X$ within the conventional range $[0.5, 2.0]$.
    The orange band is obtained by excluding the blue band region that is incompatible with the LQCD data in the region $\mu = 1.0$--$1.4\,\text{GeV}$, and $X$ is limited to the range $[1.2, 2.0]$.}
    \label{fig:pres}
\end{figure}

Here, we compare the weak-coupling results to the lattice results in QCD$_I$ in the continuum limit~\cite{Abbott:2024vhj}.
In Fig.~\ref{fig:pres}, we show the LQCD data with 1$\sigma$ error by the gray band, and the weak-coupling result $P_{\rm pQCD} + \delta P$ by the blue band;
they have a substantial overlap with each other.
We stress that the consistency between the weak-coupling and LQCD results is achieved without any fine-tuning, and the only ambiguity in the calculation is the choice of $\Lambdabar$.
The blue uncertainty band for $P_{\rm pQCD} + \delta P$ is evaluated by varying $\Lambdabar$ between $\mu < \Lambdabar < 4\mu$ around a central value $2\mu$, which is a typical scale of the system.

One goal of this work is to verify the applicable limit of weak-coupling results, in particular $\Delta$.
We claim Fig.~\ref{fig:pres} is empirical evidence for the weak-coupling formulas (\ref{eq:ppqcd}-\ref{eq:pcond}) to be valid down to $\mu \sim 0.8\,\GeV$ although it has been believed that the weak-coupling calculation of $\Delta$ is valid only at very large $\mu$ where $g = \sqrt{4\pi \alpha_s} < 1$~\cite{Rajagopal:2000rs}.
We list below the supporting evidence for the validity of the weak-coupling formulas down to $\mu = 0.8\,\GeV$.
The convergence of the perturbative series is generally good in finite-$\mu$ QCD, especially when contrasted to the finite-temperature QCD.
Also, the derivation of the gap equation is valid as long as the scale separation $\Delta \ll m_{\rm D} \ll \mu$ is fulfilled, where $m_{\rm D}^2 = 2 \alpha_s \boldsymbol{\mu}^2 / \pi$ is the gluon Debye mass.
As expected, the scale separation is clearly seen down to $\mu \sim 0.8\,\GeV$ in Fig.~\ref{fig:gap}.
This lower limit of the applicability is also consistent with high-energy QCD phenomenology, in which the lower limit of the applicability of pQCD is usually $\Lambdabar \sim 1\,\GeV$.
For the rest of the paper, we set the lower limit of applicability to $\mu = 0.8\,\GeV$ (vertical dashed line in Figs.~\ref{fig:pres}, \ref{fig:gap}).

\section{New prescription for setting $\Lambdabar$}

The only ambiguity in the weak-coupling formulas (\ref{eq:ppqcd}-\ref{eq:pcond}) is in the choice of $\Lambdabar$.
As explained above, this uncertainty is conventionally estimated by varying $\Lambdabar$ around a central value $2\mu$ by a factor of two.
The range and the central value are chosen based on historical practice, so this ad-hoc scale-variation procedure may not quantify the actual size of the error.
Recently, several authors have attempted to properly characterize the missing higher-order uncertainty in perturbative calculations and distinguish it from the ad-hoc scale variation error using statistical methods~\cite{Cacciari:2011ze, Forte:2013mda, Bagnaschi:2014wea, Bonvini:2020xeo, Duhr:2021mfd, Gorda:2023usm} (see also Ref.~\cite{Wu:2013ei} and reference therein).

Here, we discuss an alternative prescription for the choice of $\Lambdabar$ using the non-perturbative physical quantities obtained from the LQCD$_I$ computation.
Since a perturbative series is at best asymptotic series, we can view it as a workable approximation of the physical quantities nonperturbatively defined on the lattice.
Then, we determine the range of $X=\Lambdabar/(2\mu)$ so that the weak-coupling results $P_{\rm pQCD} +\delta P$~(\ref{eq:ppqcd}, \ref{eq:pcond}) is consistent with the uncertainty band of the LQCD$_I$ computation.
As shown in Fig.~\ref{fig:pres}, the majority of the blue band with the conventional scale setting $X \in [0.5, 2]$ is inconsistent with the LQCD data.
By performing a nonlinear least-squares fit of $P_{\rm pQCD} + \delta P$ to the upper bound of the LQCD data, we exclude the inconsistent part of the blue band and obtain the optimal range $X \in [1.22, 2]$.
The fit is performed in the intermediate window $\mu \in [1.0, 1.4]\,\GeV$ since the weak-coupling and the LQCD$_I$ results become more unreliable as $\mu$ become smaller and larger, respectively.
We note that we keep the upper bound of $X$ the same since it is unlikely on the physical ground that the interaction scale is even larger than $4\mu$.

Because $\alpha_s$ is the same for QCD$_I$ and QCD$_B$, the range of $X$ obtained here applies to both theories.
One can use the same lattice-informed value of $X$ in QCD$_B$ as long as the formulas~(\ref{eq:ppqcd}-\ref{eq:pcond}) at this specific order of expansion are concerned, and thus the uncertainty of $P_{\rm pQCD}$, $\Delta$, and $\delta P$ in QCD$_B$ shrinks.

\section{Implication to color superconductivity}

\begin{figure}
    \centering
    \includegraphics[width=0.95\columnwidth]{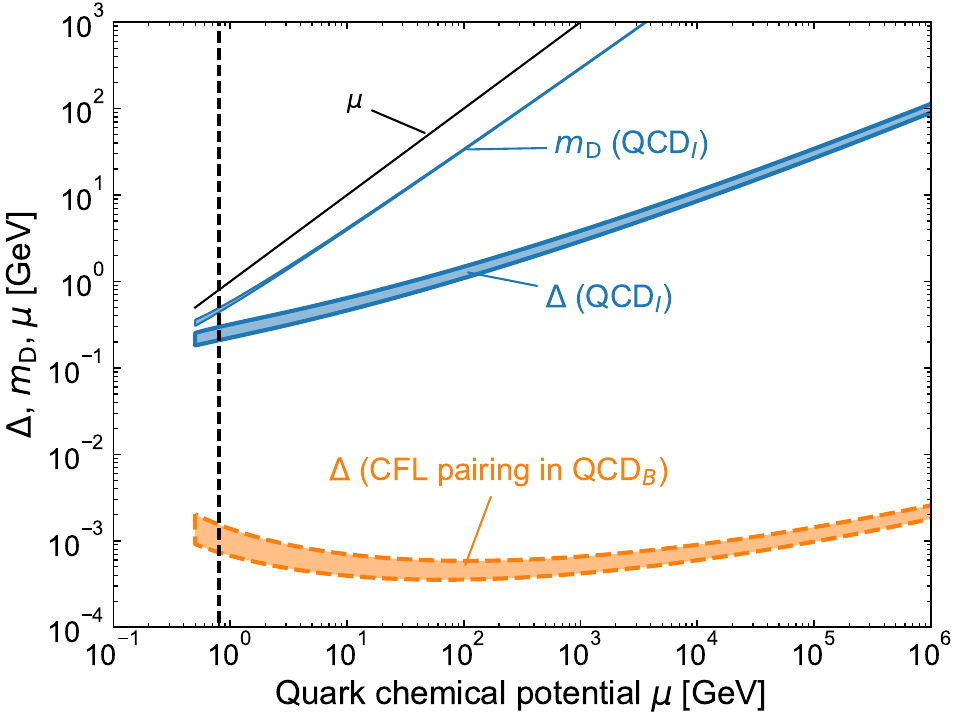}
    \caption{Magnitudes of the gap $\Delta$ in QCD$_I$ and the CFL phase of QCD$_B$, and the Debye mass $m_{\rm D}$.  The error bands are obtained by using the scale parameter $X \in [1.22, 2]$ from the LQCD fit. }
    \label{fig:gap}
\end{figure}
To date, the magnitude of the color-superconducting gap $\Delta$ in QCD$_B$ is still under debate~\cite{Baym:2017whm, Leonhardt:2019fua, Braun:2021uua, Kurkela:2024xfh, Geissel:2024nmx} (see also \cite{Nagata:2018mkb, Ito:2020mys, Yokota:2023osv}).
Here, we show $\Delta$ obtained in the weak-coupling regime~\eqref{eq:gap}, which is verified to be valid against the LQCD data above.
In Fig.~\ref{fig:gap}, we show $\Delta$ in QCD$_B$ and QCD$_I$~\eqref{eq:gap}.
While $\Delta$ in QCD$_I$ is large, $\Delta$ in the color-flavor locked (CFL) phase of QCD$_B$ is orders of magnitude smaller.
This can be understood from the difference in the strength of the one-gluon exchange attraction~\cite{Son:2000xc}, which is accounted for by the factor $c_R$ in Eq.~\eqref{eq:gap}.

The size of the CFL gap in the weak-coupling regime at $\mu = 0.8\,\GeV$ is $\Delta_{\mathrm{CFL}} \simeq 0.7\mbox{--}1.5\,\MeV$, where the range corresponds to the range of $X$.
This is consistent with the bound $\Delta_{\mathrm{CFL}}(\mu = 0.87\,\GeV) \leq 216\,\MeV$ recently obtained from the NS observables~\cite{Kurkela:2024xfh}.
The contribution of $\delta P$ from the weak-coupling $\Delta_{\rm CFL}$ to the bulk EoS is suppressed by the factor $\delta P / \Pid \sim (\Delta_{\rm CFL} / \mu)^2$ relative to $P_{\rm pQCD}$~\eqref{eq:ppqcd};
it is less than $\sim 0.01\%$ correction at $\mu = 0.8\,\GeV$.
Therefore, in the weak-coupling regime, the color-superconductivity effect on the bulk EoS is totally negligible.
This justifies the pQCD input used in the NS phenomenology, in which the contribution of the color-superconducting gap is usually neglected.

The weak-coupling calculation of $\Delta_{\rm CFL}$ may also modify the QCD phase diagram in the high-$\mu$ region.
In strange quark matter, the strange quark mass $m_s$ stresses the CFL pairing by splitting the Fermi momenta of different quark flavors apart~\cite{Clogston:1962zz,1962ApPhL...1....7C}.
The CFL vacuum remains favored over the unpaired vacuum if $\Delta_{\rm CFL}$ is larger than the stress induced by $m_s$, i.e., $\Delta_{\rm CFL} \gtrsim m_s^2 / (4\mu)$~\cite{Rajagopal:2000ff,Alford:2001zr}.
At $\mu = 0.8\,\GeV$, the condition $\Delta_{\rm CFL} \gtrsim m_s^2 / (4\mu)$ is not satisfied for the weak-coupling $\Delta_{\rm CFL}$.
The size of the CFL gap becomes comparable with the stress $m_s^2 / (4\mu)$ only at $\mu \gtrsim  1.4$--$2.4\,\GeV$, where the range corresponds to the range of $X$.
We took into account the running of $m_s(\Lambdabar)$, and fixed the value $m_s(2\,\GeV) = 93.4\,\MeV$~\cite{ParticleDataGroup:2022pth}.
This suggests that even in the weak-coupling regime, the CFL phase may not be the ground state below $\mu_B \sim 4\mbox{--}7\,\text{GeV}$.

We stress that this does not exclude the possibility of large $\Delta$ and the existence of the CFL phase in the core of NSs, inside which $\mu_B$ ranges only up to $\mu_B \simeq 1.7\,\GeV$.
This was indeed the original intention of the color superconductivity, namely, a small gap in the weakly-coupled regime is enhanced by instanton-induced interactions~\cite{Alford:1997zt, Rapp:1997zu}.
See also Ref.~\cite{Schafer:2004yx} for the instanton contribution in the dilute gas approximation.
Because the NS mass-radius constraint can only be sensitive to the large value of $\Delta$, whether $\Delta$ is large or not inside NSs has to be verified by other means more sensitive to $\Delta$, e.g., the transport properties.

\section{Impact on EoS of NS matter}
\begin{figure}
    \centering
    \includegraphics[width=0.95\columnwidth]{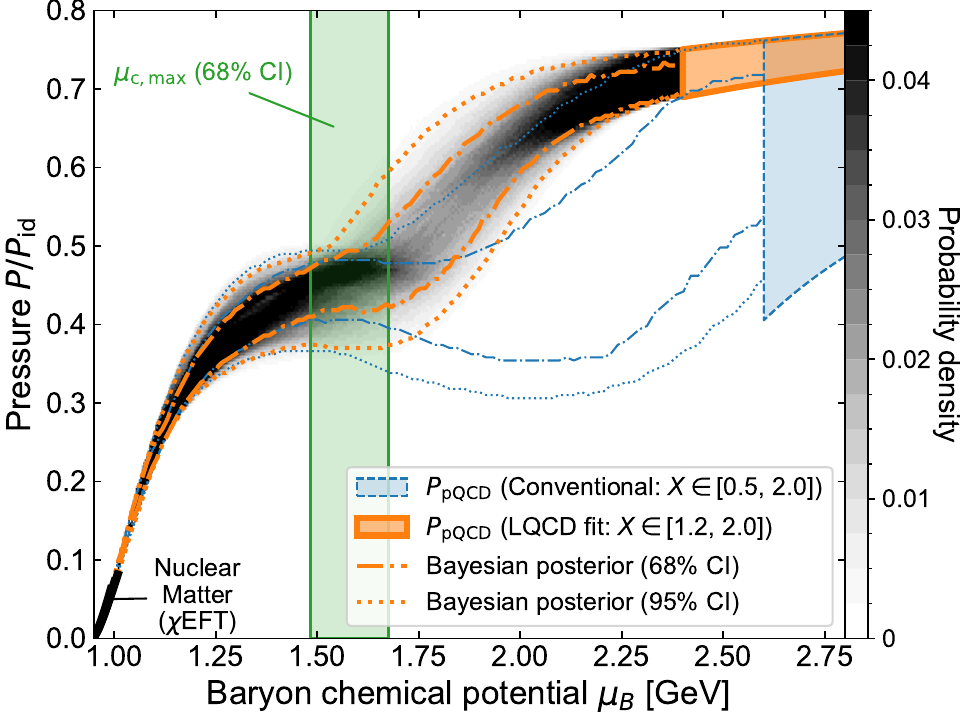}
    \caption{The posterior distribution of the Bayesian inference of NS EoS incorporating the weak-coupling inputs, i.e., requiring every EoS to connect to the weak-coupling values at $\muB = \mu_{\rm pQCD}$. 
    The blue and orange lines correspond to the conventional and LQCD-improved weak-coupling inputs, respectively.}
    \label{fig:bayes}
\end{figure}

The fitting prescription for $X$ above shrinks the relative range of $P/ P_{\rm id}$ from $\pm 31\%$ (blue shaded band in Fig.~\ref{fig:bayes}) to $\pm 4.2\%$ (orange shaded band in Fig.~\ref{fig:bayes}).
Also, from the discussion above, we slightly extend the applicable limit of the pQCD input down to $\mu_B=2.4\,\GeV$ from the customary choice $\mu_B = 2.6\,\GeV$.
Here, we discuss its impact on the NS EoS.
To this end, we perform the Bayesian inference incorporating the weak-coupling QCD inputs.
As discussed above, we can safely neglect the contribution of $\delta P$.
We follow the constant-likelihood approach developed in Refs.~\cite{Altiparmak:2022bke, Jiang:2022tps}.

In Fig.~\ref{fig:bayes}, we plot the 68\% credible interval (CI) and 95\% CI of the posterior by dash-dotted curves and dotted curves, respectively.
The conventional $X$ is marked by thin blue curves and the fitted $X$ is marked by thick orange curves.
One can see that the effect of the reduction in $X$ already appears around $\mu_{\rm c, \max}$, which is $\muB$ at the center of the maximum-mass NSs.
Limiting the range of $X$ excludes the region of the EoS with smaller values of $P/P_{\rm id}$ around $\mu_{\rm c,\max}$;  this implies that the softer EoS is favored over the stiffer EoS at higher density.

To quantitatively discuss the stiffness of the EoS, we consider the behavior of the trace anomaly $\Theta \equiv  \varepsilon - 3P$~\cite{Fujimoto:2022ohj}.
When $\Theta$ is positive, the softer EoS is favored at high density.
One can compute the Bayes factor $\calB_{H_0}^{H_1}$, which quantifies the extent to which the observed data $D$ support some given hypothesis $H_1$ over a hypothesis $H_0$:
\begin{equation}
    \calB_{H_0}^{H_1} = \frac{\Pr(D|H_1)}{\Pr(D|H_0)} = \frac{\Pr(H_1 |D) / \Pr(H_0 |D)}{\Pr(H_1)/\Pr(H_0)}\,.
\end{equation}
For the conventional $X$, $\calB^{\Theta \geq 0}_{\Theta < 0} = 3.53$, and for the fitted $X$, it is increased to $20.3$; it gives strong evidence for the EoS to have the positive $\Theta$ at high density~\cite{jeffreys1998theory}.
This strongly implies the EoS in the NS core is soft.

We note that what we have shown here is the most optimistic scenario in which the pQCD input affects the inferred result as large as possible~\cite{Komoltsev:2023zor}.
This treatment suffices here as the purpose of this Bayesian analysis is to show the maximum impact of the new prescription for fixing $\Lambdabar$.

\section{Conclusions and outlook}

We considered synergy between first-principles calculations of QCD, namely, perturbation theory and lattice simulation.
LQCD information can be synthesized into the weak-coupling calculation by tightening the range of renormalization scale $\Lambdabar$; this can further constrain the NS EoS.
We also point out that the CFL phase may not be the ground state up to $\mu_B \sim 4\,\text{GeV}$ in the weak-coupling regime.
Perturbation theory will allow a precise determination of the QCD phase diagram down to $\muB\sim2.4\,\GeV$.

\begin{figure}
    \centering
    \includegraphics[width=0.95\columnwidth]{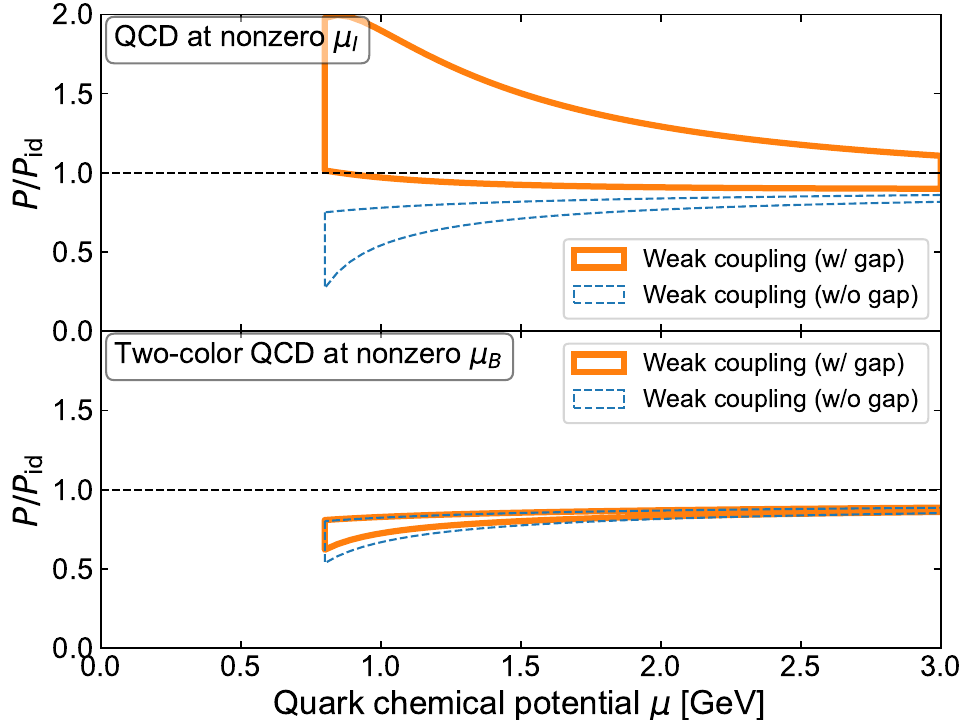}
    \caption{Comparison of the weak-coupling behavior of $P/P_{\rm id}$ in the systems in which LQCD simulations are feasible, i.e., QCD$_I$ (top) and QC$_2$D (bottom).}
    \label{fig:compare}
\end{figure}

Further test of the weak-coupling formulas~(\ref{eq:ppqcd}-\ref{eq:pcond}) can be provided in SU(2) gauge theory at finite $\muB$ (QC$_2$D), another sign-problem-free theory~\cite{Hands:2006ve, Cotter:2012mb, Boz:2019enj, Begun:2022bxj, Iida:2022hyy, Iida:2024irv}.
The gap in QC$_2$D turns out to be small because the color factor in Eqs.~(\ref{eq:gap}, \ref{eq:pcond}) has a smaller value $c_R = 3/4$. Also, the coupling constant $\alpha_s \propto 1/\beta_0 \propto 1/N_c$ in QC$_2$D is larger relative to QCD in $N_c=3$, which results in the further parametric suppression of $\Delta$ by the factor $g^{-5}$.
This leads to a sharp contrast in the behavior of weak-coupling EoSs in QCD$_I$ and QC$_2$D as shown in Fig.~\ref{fig:compare}.
We note that $\Lambda_{\overline{\mathrm{MS}}}$ in QC$_2$D has to be determined independently from the lattice calculation, and here we fix it tentatively to $\Lambda_{\overline{\mathrm{MS}}} = 300 \,\MeV$.
If the lattice calculation in QC$_2$D gives a smaller value of $P/P_{\rm id}$, this will be a further strong indication for the validity of the weak-coupling results.

\begin{acknowledgments}
The author gratefully acknowledges
William Detmold,
Tetsuo Hatsuda,
Sanjay Reddy,
Farid Salazar,
and
Thomas Sch\"afer
for useful comments and discussions.
The author would especially like to thank Risto Paatelainen, Kaapo Sepp\"anen, and Aleksi Vuorinen for pointing out the issues related to the number of flavors.
The author thanks the Yukawa Institute for Theoretical Physics at Kyoto University and RIKEN iTHEMS, where part of this work was completed during the International Molecule-type Workshop ``Condensed Matter Physics of QCD 2024'' (YITP-T-23-05).
The author is supported by the Institute for Nuclear Theory's U.S.\ DOE Grant No.\ DE-FG02-00ER41132.

\end{acknowledgments}

\appendix

\section{Classification of various corrections to the weak-coupling formula}
\label{sec:corr}
Here, we argue that $\delta P$ is the only large correction in the weak-coupling regime of QCD$_I$.
One can classify possible corrections to $P_{\rm pQCD} \sim \mu^4$ by expanding $P$ in terms of $\mu$~\cite{Alford:2004pf}
\begin{equation}
    P = P_{\rm pQCD} + a_2 \mu^2 + a_0\,,
    \label{eq:power}
\end{equation}
where the odd powers of $\mu$ are absent because of $C$-symmetry in QCD.
The coefficients $a_2$ and $a_0$ are treated approximately as constants since their $\mu$-dependences are weak.
Aside from $\delta P$, the corrections of the quadratic order are a current mass of flavor-$f$ quarks $a_2 \propto m_f^2$, and the temperature correction $a_2 \propto T^2$ as the lattice data is measured at $T \sim 20\,\MeV$.
Both of these corrections are less than $ 1\%$ relative to $P_{\rm pQCD}$ at $\mu = 0.8\,\GeV$.
The constant correction term $a_0$ may originate from the bag constant $a_0 = -B$ and instantons~\cite{Shuryak:1978yk, *Shuryak:1978ew, *Shuryak:1982hk, Kallman:1979wm, Abrikosov:1980nx, *Abrikosov:1981qb, *Abrikosov:1982mi, *Abrikosov:1983hxi, AragaodeCarvalho:1980de, Chemtob:1980tu, Baluni:1980db}.
The bag constant of a typical size $B^{1/4} \sim 150$--$200\,\MeV$ amounts to $\sim 2$--7\% correction at $\mu=0.8~\GeV$ and its contribution decreases by a factor $\mu^{-4}$ with increasing $\mu$.
The instanton contribution is small as it is suppressed by quark masses $\propto \prod_f (m_f / \Lambda_{\overline{\rm MS}})$.
From the above consideration, we conclude that $\delta P$ is the only largest correction to $P_{\rm pQCD}$.

\section{Details of the Bayesian analysis}
Here, we outline the method of our Bayesian analysis.
We follow the constant-likelihood approach developed in Refs.~\cite{Altiparmak:2022bke, Jiang:2022tps}, which can be regarded as a Bayesian extension of Ref.~\cite{Annala:2019puf}.

For the prior distribution, we randomly sample EoSs by modeling them with the piecewise sound-speed parametrization introduced in Ref.~\cite{Annala:2019puf}.
We use the crust EoS at low $\mu_B$~\cite{Baym:1971pw}, and use the EoS from the $\chi$EFT~\cite{Drischler:2020fvz} up to $\mu_B = \mu_0$, where $\mu_0 \simeq 1.0\,\GeV$.
The density and pressure at $\mu_0$ are $n_0 \simeq 0.24\,\fm^{-3}$ and $P_0 \simeq 8.8_{-1.5}^{+1.4}\,\MeV/\fm^3$, respectively.
The error corresponds to $\pm 1 \sigma$ credible region presented in Ref.~\cite{Drischler:2020fvz}, and we sample equal numbers of the EoS at average and at $\pm 1 \sigma$ credible interval (CI) in the prior distribution.
We randomly sample parameters $\mu_i$ and $v_{s,i}^2$ ($i=1,\ldots,N$) from the uniform distributions $U(\mu_0, \mu_{\rm pQCD})$ and $U(0, 1)$, respectively.
We choose the number of segments as $N = 5$, and $\mu_{\mathrm{pQCD}} = 2.6\,\GeV$ for the conventional X range $[0.5, 2]$ and $\mu_{\mathrm{pQCD}} = 2.4\,\GeV$ for the fitted X range $[1.2, 2]$.
Then, we linearly interpolate between each segment boundary $(\mu_i, v_{s,i}^2)$, and from this linearly-interpolated $v_s^2(\mu)$, one can construct the complete thermodynamic function $P(\mu)$ through the integration:
\begin{equation}
    P(\mu) = P_0 + n_0 \int_{\mu_0}^{\mu} d\mu' \exp\left[\int_{\mu_0}^{\mu'} \frac{d\mu''}{\mu'' v_s^2(\mu'')}\right]\,.
\end{equation}

For the Bayesian update step, we incorporate the pQCD input as a likelihood, i.e., we require an EoS to be $P_{\min} < P(\mu_{\rm pQCD}) < P_{\max}$, where $P_{\min}$ and $P_{\max}$ are $P_{\rm pQCD}$ evaluated for the minimum and maximum $X$, respectively.
We also use the observations of $2M_\odot$ radio pulsars~\cite{Demorest:2010bx, Antoniadis:2013pzd, NANOGrav:2019jur}, $R$ measurements from NICER~\cite{Riley:2019yda, Miller:2019cac, Miller:2021qha, Riley:2021pdl}, and the binary tidal deformability $\tilde{\Lambda}$ measurements from the GW170817 event~\cite{LIGOScientific:2017vwq, LIGOScientific:2018cki}.
Namely, following the procedure in  Refs.~\cite{Altiparmak:2022bke, Jiang:2022tps}, we impose the following conditions on an EoS:
$M_{\rm max} > 2~M_\odot$,
$R(M = 2.0~M_\odot) > 10.75~\text{km}$,
$R(M = 1.1~M_\odot) > 10.80~\text{km}$, and $\tilde{\Lambda} < 720$ at a chirp mass $\mathcal{M}_{\rm chirp} = (M_1 M_2)^{3/5} (M_1 + M_2)^{-1/5} = 1.186 M_\odot$ for any choice of mass in the binary $M_{1,2}$ with a mass ratio $M_2 / M_1 > 0.73$.

The probability density shown in Fig.~\ref{fig:bayes} is estimated using the histogram of $4.1\times 10^4$ posterior sample EoSs in the $491 \times 500$ equal-width bins on the $(\muB, P/P_{\rm id})$-plane within the range $[926\,\MeV, \mu_{\rm pQCD}] \times [0,2]$.
The probability density is normalized by the total number of posterior samples.

\bibliography{scale}

\end{document}